\documentclass[table]{ccjnl}
\usepackage{lipsum,amsmath}
\usepackage{cuted}
\usepackage{bm}
\usepackage{microtype}
\usepackage{xcolor}
\usepackage{gbt7714}
\graphicspath{{figures/}}

\title{From Earth to Space: A First Deployment of 5G Core Network on Satellite}
\author{Ruolin Xing\inst{1}, Xiao Ma\inst{1}, Ao Zhou\inst{1}, Schahram Dustdar\inst{2}, Shangguang Wang\inst{1, *}\corinfo{sgwang@bupt.edu.cn}}
\receiveddate{TBD}
\reviseddate{TBD}
\Editor{TBD}

\address[1]{State Key Laboratory of Networking and Switching Technology, Beijing University of Posts and Telecommunications, Beijing 100876, China}
\address[2]{Distributed Systems Group, TU Wien, 1040 Vienna, Austria}

\begin{document}

\maketitle

\begin{abstract}
Recent developments in the aerospace industry have led to a dramatic reduction in the manufacturing and launch costs of low Earth orbit satellites. The new trend enables the paradigm shift of satellite-terrestrial integrated networks with global coverage. In particular, the integration of 5G communication systems and satellites has the potential to restructure next-generation mobile networks. By leveraging the network function virtualization and network slicing, the orbital 5G core networks will facilitate the coordination and management of network functions in satellite-terrestrial integrated networks. We are the first to deploy a lightweight 5G core network on a real-world satellite to investigate its feasibility. We conducted experiments to validate the onboard 5G core network functions. The validated procedures include registration and session setup procedures. The results show that the 5G core network can function normally and generate correct signaling.
\keywords{5G Core Network; Satellite Communications; Satellite Internet}
\end{abstract}

\section{Introduction}
\label{Introduction}
Traditional terrestrial communication systems have flourished in many ways. Especially, the 5G networks have enabled much higher data rate, ultra low latency, and massive network capacity \cite{3gpp.23.501}. However, there are still some inherent disadvantages that are difficult to resolve \cite{giambene2018satellite}. Mobile networks have poor coverage in remote areas. They suffer long-distance transmission delays. Besides, vulnerability to natural disasters makes terrestrial communication systems unavailable in extreme cases. Meanwhile, the advancements in aerospace technology have brought the satellite industry a resurgence, from serving vertical fields to providing more universal services. Previously, operators provided limited communication services to dedicated users through geostationary satellites. Recent projects like Starlink have used low Earth orbit (LEO) constellations to deliver global broadband Internet access.The urgent needs and the rapid developments boost the integration of satellites and 5G systems. In addition, the significance of incorporating satellites is not merely enhancing terrestrial networks. The dominant class of services is content delivering, which takes about 90\% of the total traffic in telecommunication networks. The satellites will have better cost effectiveness in broadcasting/multicasting connectivity and makes the services more convenient.

There have been numerous efforts devoted to the integration of 5G and satellites \cite{boero2018satellite}. The 3rd generation partnership project (3GPP) initiated activities towards non-terrestrial networks that study the role of satellites in 5G and acknowledge long-term research within B5G and 6G \cite{3gppradio, 3gpp.38.811, lin20215g}. Besides, prior works \cite{wang2019convergence, kodheli2020satellite} investigated the standardization of non-terrestrial networks. Reference \cite{bacco2019networking} further summarized the networking challenges for non-terrestrial networks in the 5G ecosystem. Studies \cite{kodheli2017integration, gopal2018framework} focused on the satellites which were used as a complement for radio access networks in 5G communication systems. Technology companies like Lynk, OmniSpace, Vulcan Wireless, Lockheed Martin, and AST SpaceMobile have already combined 5G and satellite to enable direct connections between user equipments and satellites. While researchers explored the advantages brought by the satellites being a part of transmission networks \cite{abdelsalam2019implementation, bujari2020virtual}. In addition, satellite networks can bring extra use cases for the 5G network slicing, which has been studied in \cite{bisio2019network, lei2021dynamic}.

5G core networks have evolved into software-defined communication systems based on virtualization technology \cite{3gpp.23.501}. With the explosive growth in scale and business, the management, operation, and maintenance of satellite networks will be more complex. Benefited from software-defined networking and network function virtualization, 5G core networks can achieve distributed deployment and elastic scaling\cite{gupta2015survey}. Migrating some functions of 5G core networks to satellites can connect users and data networks more flexibly. With these techniques, 5G core networks have the potential to be used to implement a satellite-terrestrial integrated network\cite{gopal2018framework}. There are also studies on the role of core networks in the 6G era \cite{giordani2020non, kota20216g}. Some core network functions can be moved onboard to gain more flexibility and efficiency \cite{wang2022holistic}. In-orbit core networks may suffer from signaling storms and the data session migrations. To address these challenges, \cite{li2022case} has proposed a prototype of in-orbit core networks, which adopted a stateless design.

This paper aims to study the convergence of satellites and 5G core networks. In this direction, we introduce the motivations for deploying the 5G core network functions onto satellites. Next, we provide an overview of the satellite capabilities with onboard 5G core networks. We then propose an architecture of a satellite-terrestrial integrated network. The architecture shows the advantages of the onboard 5G core network. Finally, we present the deployment and experiment procedures of the 5G core network on the satellite. It is the first attempt to setup a satellite-terrestrial integrated network with the 5G core network deployed on a real-world satellite.

\begin{figure*}
\centerline{\includegraphics[width=36pc]{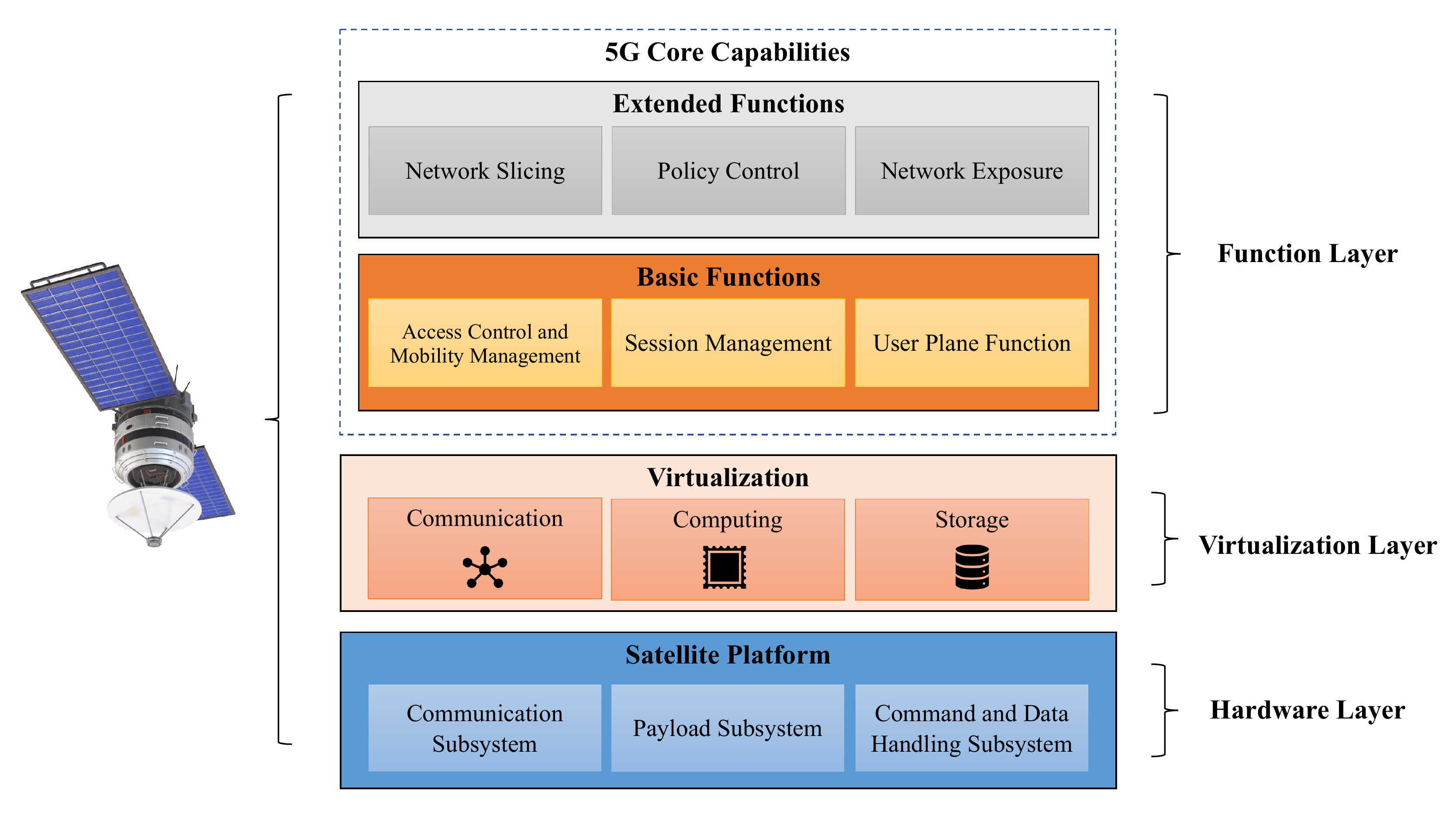}}
\caption{The capability structure of satellites with orbital 5G core networks.}
\end{figure*}

\section{Motivations}
\subsection{Main Aspects}
\paragraph{Demand Perspective} The costs of launching and manufacturing satellites continue to decrease. The LEO constellations are expected to expand exponentially in the near future \cite{denby2020orbital}. Onboard computing, communication, and storage resources will also usher in a period of explosive growth. Based on that, deploying core networks on satellites becomes practical. Meanwhile, traditional satellite services like remote sensing, navigation, and communications have great needs for computing power. The demand for in-orbit computing will drive the deployment of additional satellite services. With the 5G core networks onboard, mobile users can access the satellite services more conveniently.

\paragraph{Performance Perspective} The terrestrial mobile core networks tend to sink to the edge. The purpose is to reduce the transmission delay and improve the quality of user experience. The onboard core networks will benefit from the low-latency and wide-coverage links of LEO satellites. The terrestrial fiber paths are generally long-winded, in which the light travels at roughly \textit{2c/3} \cite{bozkurt2017internet}. While most of LEO satellites are orbiting at 500 km to 1,000 km above the Earth's surface. Previous study \cite{bhattacherjee2018gearing} showed that the LEO satellite constellation can achieve a 50\% improvement in latency over today's terrestrial networks. For typical end-to-end connections, even the relatively small constellation can (almost always) achieve latencies better than the best possible with fiber. Thus, the onboard core networks can get a considerable performance gain compared to the terrestrial ones. Additionally, onboard core networks eliminate unnecessary backhauling which takes a large fraction of the network latency. As a result, onboard core networks will reduce the control plane signaling interaction delay and speed up the user access procedures.

\paragraph{Functional Perspective} According to the 3GPP technical report, satellites will have base station functions and be a part of access networks\cite{3gpp.38.811, liu2018space}.
Access networks rely on core networks to provide mobility management, session management, and other functions.
Based on that, the onboard 5G core networks can benefit in two ways.
Firstly, the onboard 5G core networks will make users be able to get the satellite services without backhauling.
Secondly, the satellites with base station fuctions can be managed more conveniently by the onboard rather than the terrestrial core networks.
Due to the continuous change in relative position of LEO satellites, the connections between satellites and ground stations are usually unavailable.
It means that satellites are out of the control of the core network most of the time. 
Onboard core networks can integrate with future access networks which are made up of large-scale LEO satellite constellations.

\subsection{Use Cases}
\paragraph{Orbital Edge Computing} Edge computing on LEO satellites is gaining popularity recently \cite{denby2020orbital, bhattacherjee2020orbit}. Traditional satellites generally adopt the bent-pipe architecture which does no modifications to the downloaded data. The alternative is using the onboard payloads to do specific processing tasks. Typical applications include remote sensing, in-orbit AI inferencing, and space storage. The onboard core network can help to realize in-orbit edge computing services for mobile users. Additionally, the inter-satellite links (ISLs) can reduce the dependence of constellations on ground stations. Co-orbiting satellites can achieve large-scale and uninterrupted services through collaboration. Deploying core networks in such constellations can ensure higher availability of edge computing services.

\paragraph{Emergency Communication} The combination of satellite and 5G core network provides a better option for emergency communications. On the one hand, satellite communication can solve the problem of emergency communication when the ground infrastructure is damaged. The main benefits of satellite communication are wide coverage and large transmission capacity. when an emergency occurs, the satellite can act as a base station and core network facility to route traffic to any data network. On the other hand, the deployment of lightweight core networks on dedicated satellites can be regarded as an upgrade of satellite communication capabilities. The original remote sensing and experimental satellites will be transformed into satellites with emergency communication capabilities after the deployment.

\section{Overview of the onboard 5G core network}
{\setlength{\parindent}{0cm}
The renaissance of satellite networks began with reduced launch costs and pipe-lined satellite production. Large-scale commercial deployment of 5G core networks on satellites will soon be possible due to the growing resources of communication, computing, and storage. To further illustrate what a full-fledged 5G core network can bring to satellite networks, it is necessary to clarify the functionalities of different parts of satellites deployed with 5G core networks. This section provides an overview of capabilities combining satellites and 5G core networks from different layers. Figure 1 shows the structure of a typical satellite deployed with a 5G core network and reflects the organization of this section.
The 5G core network capabilities are shown in the dashed box. The solid line box named virtualization is an independent level abstracted directly on the hardware platform to provide support for functions in core networks.}

\subsection{Satellite Platform}
{\setlength{\parindent}{0cm}
A complete satellite platform consists of several subsystems.
These systems coordinate among one another to execute space missions.
Common subsystems include the electrical power system, the attitude and orbit control system, the communication system, the payload system, and the command and data handing system.
}

The subsystems manage different aspects of the satellite platform, which may be related to the onboard 5G core networks.
The communication subsystem can setup wireless links to transmit data or commands.
The data transmission links typically have higher bandwidth in the downlink.
Thus, the user plane of orbital core networks should consider corresponding design.
The payload system manages the elements that produce mission data and then relays the data back to the Earth.
The 5G core networks can be deployed in the payload subsystem.

The command and data handing system deals with telemetry and telecommand.
It provides a communication channel between the spacecraft and the ground operators.
On the uplink, it receives and decodes all telecommands and data from the communications system. These commands are then sent to the appropriate subsystem or executed directly.
On the downlink, it collects the telemetry from payloads or other subsystems, and multiplexes the data into frames for later transmission. The operations toward onboard 5G core networks are executed by the command and data handing system.

Satellite platforms also have further influences on the onboard core networks.
Firstly, the radio frequency bands of satellites are scarce resources.
As the frequency goes higher, the antenna size becomes smaller.
However, smaller antenna can be affected more easily by adverse weather conditions.
For satellite communications, frequency bands between 1-50 GHz which include L, S, C, X, Ku, K, Ka, and Q/V bands are heavily used currently.
To increase the link throughput, researchers are considering implementing terahertz and optical bands in future satellite communications \cite{tan2020thz}.

Secondly, the enhanced onboard processing capability makes satellite systems more intelligent.
This evolution mainly involves signal processing and packet processing.
The signal processing can help effective beam switching among satellite beams.
It allows satellites to provide more flexible wireless resource management, which enables fine-grained policy and quality of service control.
The packet processing can enable packet switching functionalities on satellite systems to support constellation-wide networking.

\begin{figure*}
\centerline{\includegraphics[width=40pc]{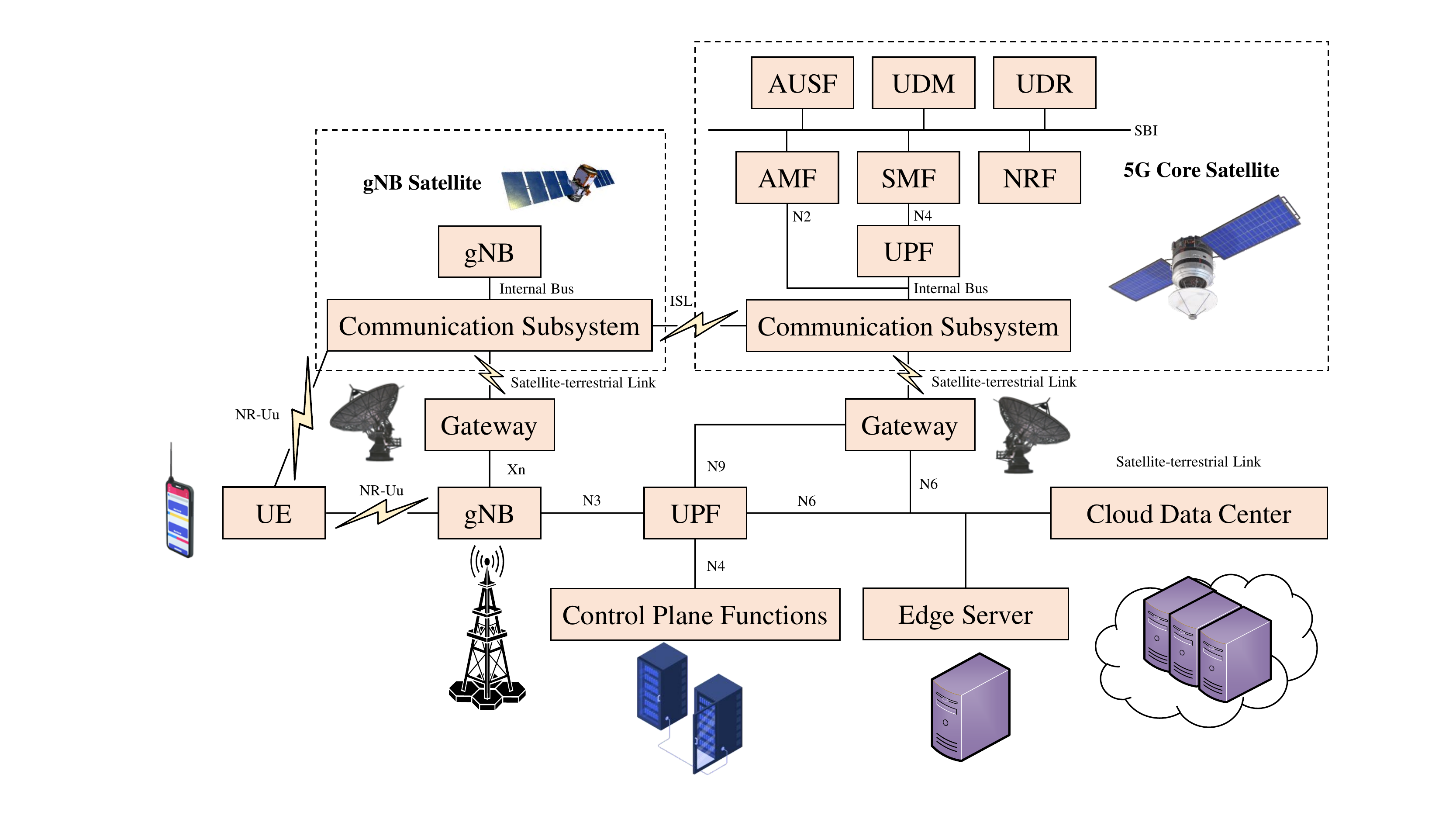}}
\caption{The architecture of 5G core networks deployed on both satellites and terrestrial data centers.}
\end{figure*}

\subsection{Virtualization}
{\setlength{\parindent}{0cm}
By leveraging the virtualization technology, software-defined wireless networks are becoming universal standards.
The satellite industry is also embracing this new trend.
Recently, the first programmable telecommunications satellite, Eutelsat Quantum \cite{esa2021} was launched successfully.
Traditional satellite networks use dedicated hardware to transmit and process data, making networks hard to manage and scale.
With the advancements in edge computing \cite{mansouri2021review}, the virtualization can be applied to resource-constrained devices. Thus, it is now applicable to tackle this problem by developing software and hardware integration platforms on satellites and operating on the virtualized resources directly.}

On one hand, the heterogeneity of satellite resources makes it difficult to be effectively used and flexibly controlled. Virtualization provides a layer of logic encapsulation on top of the underlying heterogeneous communication, computing, and storage resources. This allows satellite developers to conveniently call basic capabilities on the interfaces provided by the virtualization layer. On the other hand, the highly dynamic user demands and limited resources increase the complexity of operating satellite communication networks. It is necessary to fully unleash the potential of virtualization to achieve agile control, flexible management, and reduction of maintenance costs.

As Figure 1 shows, virtualization serves as the foundation for other functionalities of satellites deployed with 5G core networks. Based on that, different layers of the integrated network can utilize virtual resources at varying degrees. For instance, the virtualization process is supposed to cover the full end-to-end link including access networks, transmission networks, and core networks. Based on that, slicing for the integrated network and dynamic resource assignment can be realized.

Software defined networking and network function virtualization are two key enablers of 5G core networks \cite{bao2014opensan}. They also play an important part in the virtualization of satellite networks. Network function virtualization builds the programmable data plane, and provides basic networking capabilities. The control plane follows the general rules of software defined networking. But the control logic can be distributed to fit the high-dynamic characteristics of satellites. The upper management plane should be cloud-native to make full use of scarce resources on satellites. It can also choose the path of serverless and make services with the granularity of meta-function to eliminate redundancy.

\subsection{Basic 5G Core Network Functions}
{\setlength{\parindent}{0cm}
Software defined networking separates the network into the data plane and the control plane. The control plane is centralized, which allows networks to be controlled much more easily than their traditional distributed counterparts. The basic 5G core network functions include a complete set to setup data paths from mobile users to data networks inspired by software defined networking. These functions are the minimal requirements to execute the user registration and session management procedures. The three key functions are: the access control and mobility management function (AMF), the session management function (SMF), and the user plane function (UPF). There are also other imperative network functions such as functions for authentication and data persistance. The orbital network functions behave differently compared with the terrestrial network functions.}

\begin{figure*}
\centerline{\includegraphics[trim=0 3cm 0 3.5cm, width=40pc, clip]{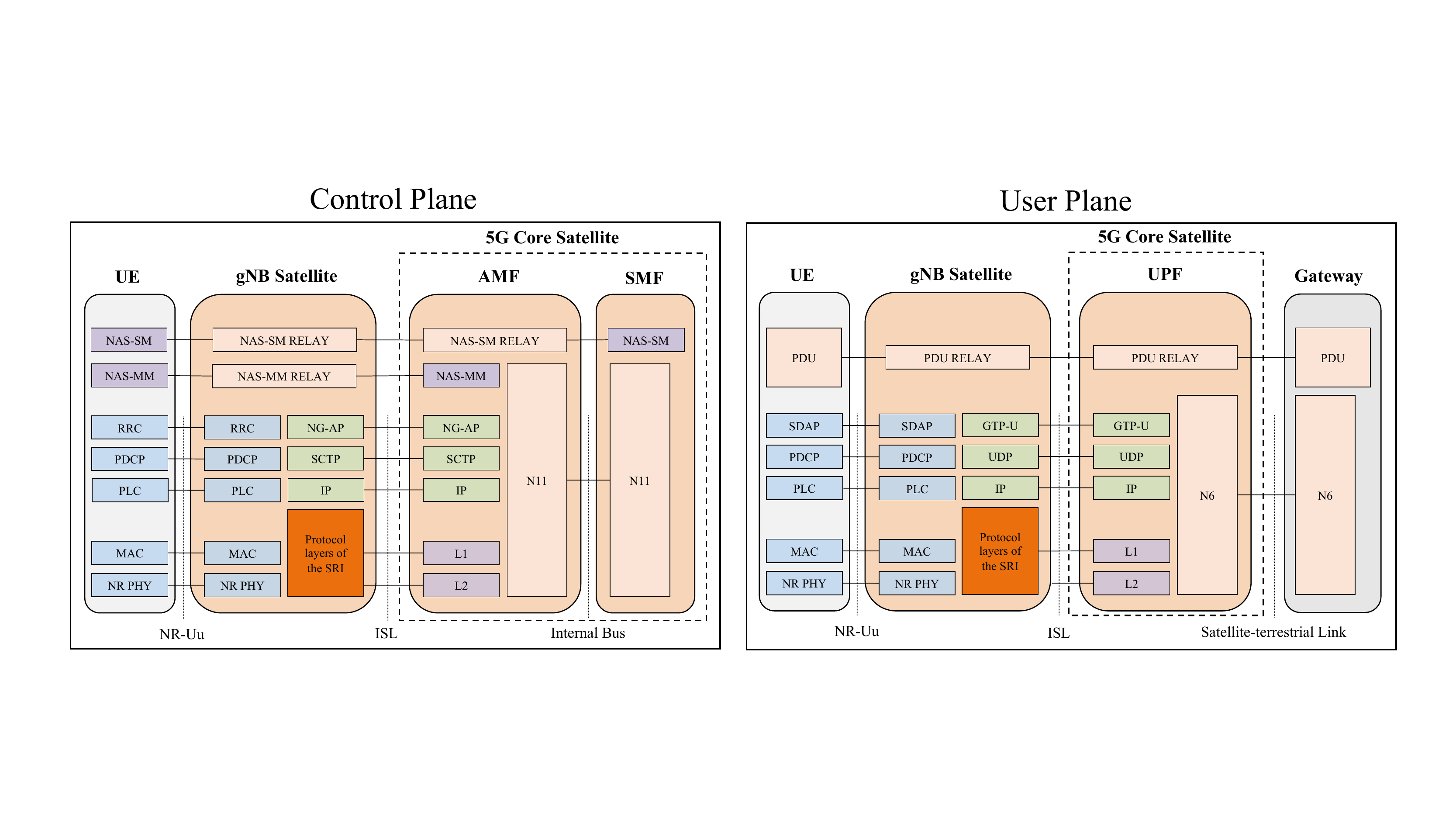}}
\caption{The protocol stacks of the onboard 5G core network in both the control plane and user plane.}
\end{figure*}

The user plane funtion represents the data plane evolution of the control and user plane separation strategy. It acts as the interconnect point between the mobile infrastructure and the data network. What's more, it's the protocol data unit session anchor point for providing mobility within and between different radio access technologies. Onboard user plane funtion can handle the packet routing and forwarding. For instance, it can perform the role of an uplink classifier which splits the traffic into different data networks. One of the use case is conducting the user traffic to the onboard services.

The relatively high-speed movement between satellites and user terminals brings new challenges. If the onboard AMF adopts same mobility management as terrestrial networks, the movement will cause signaling storms and trigger frequent registrations. In \cite{li2022case}, a state-funtion-location decouping approach is proposed to renovate AMF. This design further eliminates the needs of mobility management for the onboard AMFs. Thus, the AMFs are no longer the mobility anchors for the users. The users deal with their own mobility. The inherent challenge is how to properly handle the infrastructure mobility of the onboard AMF. Novel handover and registration mechanisms are the key for solving this problem.

Establishing connections toward data networks is one of the basic capabilities of the core network. The session management function needs to consider the requirements from users, networks, and services. Due to the highly dynamic topology and randomly fluctuating load nature of satellites, One of the challenges is how to plan and manage reliable data paths for user sessions. In addition to the data connectivity, it should also accommodate satellite-specific business scenarios.

\begin{figure*}
\centerline{\includegraphics[trim=5 60 5 60, width=40pc, clip]{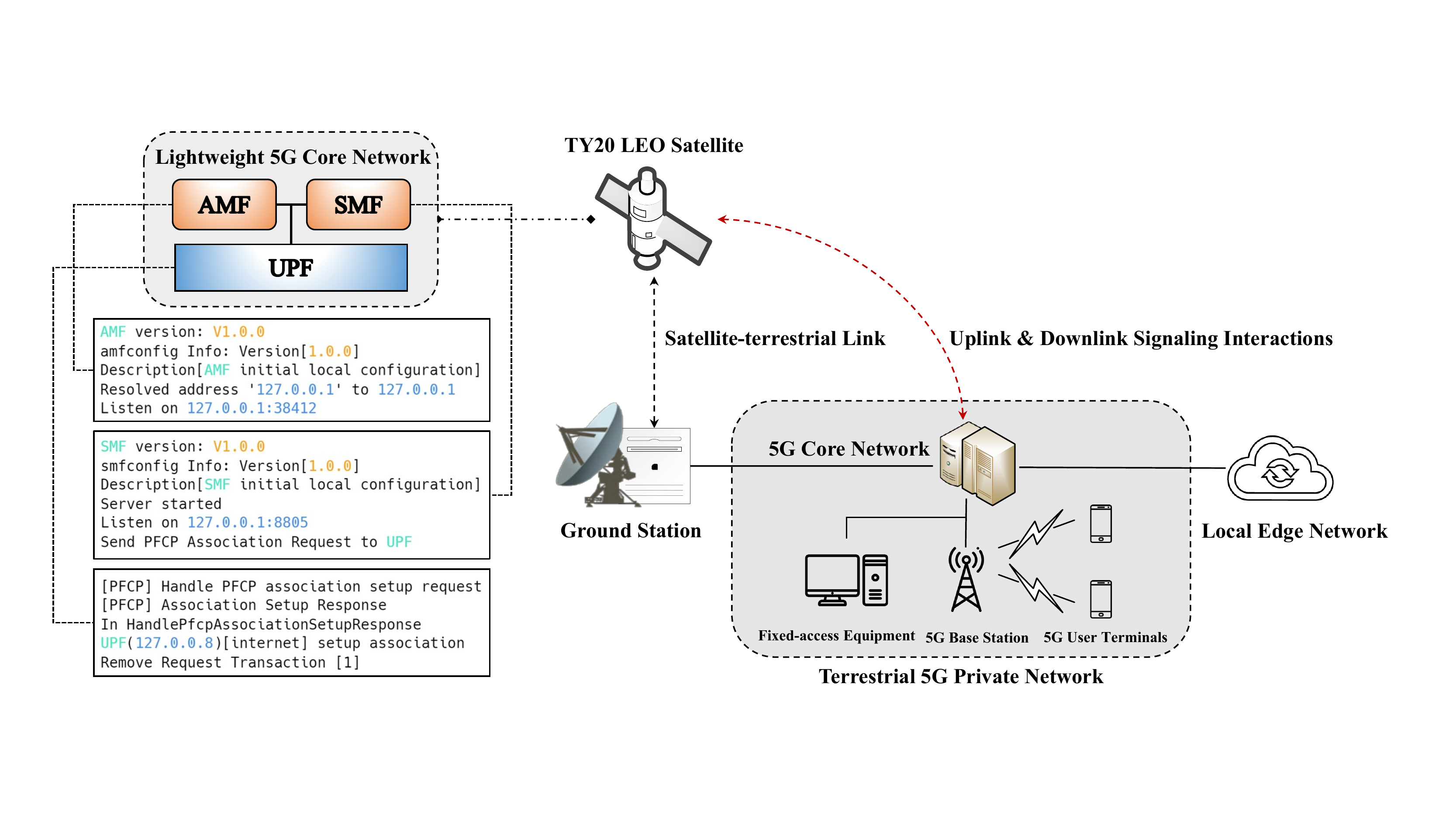}}
\caption{The deployment of the lightweight 5G core network on TY20.}
\end{figure*}

\subsection{Extended 5G Core Network Functions}
{\setlength{\parindent}{0cm}
While the basic 5G network functions aim to establish connections, the extended 5G core network functions are for more advanced features.
These features include providing differentiated capabilities for end users and business scenarios.
The virtualization technically support implementing these advanced features.
It provides standard interfaces for upper layers to manipulate the resources in a programmable way. Thus, satellites can have a certain degree of versatility and provide multiple services by using the same underlying hardware.}

The policy control function is based on the Quality of Service (QoS) guarantee over satellite-terrestrial integrated networks.
The QoS guarantee is about providing differentiated data forwarding treatment for different users, applications, and services.
By integrating business requirements of satellites and QoS capabilities, policy control can better ensure the stability and continuity of services.
The guarantee of specific data flows will become a key issue for the onboard policy control, such as supporting the intelligent remote sensing or the content delivery traffic in the same time.

Network slicing is the key technology for future satellite service provisioning.
Through the end-to-end division, networks can be customized to support dedicated services.
Network slicing can help assign network slices for long-distance and low-delay transmission, broadcast communications, or other specific scenarios in satellite networks.
However, network slices must hold with mobility of satellites.
Given that the satellite network infrastructure is highly dynamic yet predictable, we may sustain the network slices by maintaining the virtual stationarity \cite{bhattacherjee2020orbit}.

The network exposure function is for exposing the network functions and capabilities of 5G networks. Third-party enterprises can composite or customize application functions on-demand.
Thus, the network functions of the orbital 5G core networks can be fully reused.
Considering increasing demand for satellite services, network exposure function will provide multiple solutions for different types of satellite services.

\section{SYSTEM DESIGN}
{\setlength{\parindent}{0cm}
Studies of 3GPP on non-terrestrial networks define two architectures by payload type: transparent payload and regenerative payload \cite{3gpp.38.821}. If the payload is transparent, satellites will serve as repeaters that relay the signal between the user equipment and base stations. Satellites with regenerative payloads can regenerate signals received from the earth. The signal regeneration is supported by the onboard processing capability, which further makes gNB satellites and ISLs applicable.}

The 3GPP standardization activities provide basic reference for the system design. Besides, the future network demands should also be considered. For future space-air-ground integrated networks, radio access networks are expected to have a massive development by incorporating ships, high-altitude platforms, and satellites as new access methods \cite{liu2018space}. To realize ubiquitous access, some functionalities of core networks have to be split and sink to the edge to meet the requirements of time-sensitive control, autonomous management, and convenient capability exposure.

Figure 2 shows the architecture of 5G core networks deployed both on satellites and a terrestrial data center.
The radio links are marked with yellow lightning.
The communication subsystems can support the ISLs and the satellite-terrestrial links.
Due to the space limitation, please refer to \cite{3gpp.38.821} for the explanation of abbreviations.
The architecture doesn't imply a specific deployment.
For instance, the gNB functions may be deployed with the core network functions in a same satellite. By decoupling the control plane and user plane, the 5G core networks support flexible deployment. Thus, the they are independent of which orbit they are in. For instance, the 5G core deployed on LEO satellites can provide mobility management and access control, as well as simple orchestration on routing and session management. The 5G core deployed on geostationary satellites may offer service registration and discovery for constellation-wide service coordination. Figure 3 shows protocol stacks of the onboard 5G core network when a user accesses through the base station on satellites. The dotted box marks the onboard part of the core network. Due to the space limitation, please refer to \cite{3gpp.38.821} for the explanation of abbreviations. In addition, the actual interface between protocol stacks may vary under different deployment scenarios.

Satellites have limited resources and high power consumption. The 5G core network will consume higher computing power if the number of users increases. We investigate the network function pruning to reduce functional redundancy and improve efficiency. The 5G core network adopts the service-based architecture for the control plane. In a 5G core network, the smallest invokable unit is the network function service. The actual functions are implemented by operations inside the network function services. This design is convenient to ensure the relatively stable business logic of the core network. But it is not conducive to reusing the underlying functions, which will result in redundant calls between many processes. Because of scarce onboard resources and an uncertain link environment, adopting this design will make it more difficult to guarantee core network services. In essence, network function pruning of the core network consists of two steps. The first is to explore the logic issues associated with the minimal core network business procedures. And business-oriented network element services are created on that basis. In this way, the core network function streamlining can be realized.

The architecture proposed here also paves the way for future research on B5G/6G core networks. In our previous work \cite{li2021cognitive}, we have two assumptions. The first is that the 6G core network control plane would get closer to the edge to improve performance. The second is that the monolithic core network would be parted into the cloud core network and edge core network. Edge core is an autonomous network that has self-contained capabilities to deliver services and can also coordinate with other networks. The cloud core doesn't have to be full-fledged and directly connects to base stations. It's an operation and management center that organizes the edge cores to behave correctly and efficiently. There are no specific restrictions on the deployment of edge cores and cloud cores. Generally, edge cores are deployed close to various kinds of radio access networks. Cloud cores are deployed at where the cost of computing power is low.

\section{SYSTEM VALIDATION}

\begin{figure}
\centerline{\includegraphics[width=20pc]{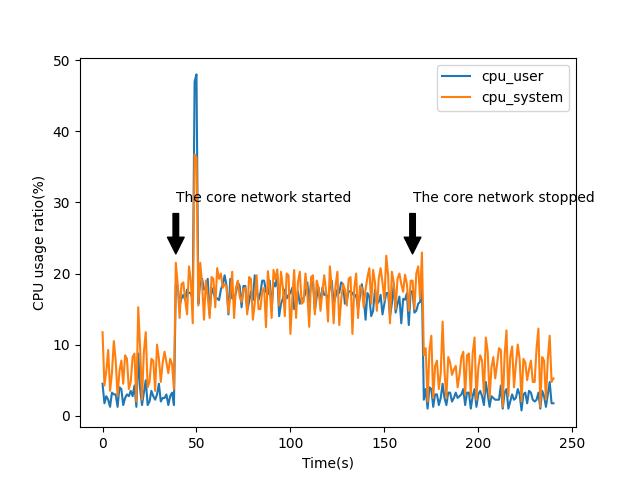}}
\caption{The CPU usage change over time.}
\end{figure}

{\setlength{\parindent}{0cm}
To verify the feasibility of orbital 5G core networks, we develop a lightweight 5G core network.
We deploy it on an in-orbit satellite, TY20.
In the experiment, we use the data transmission function in X band for the downlink and telecommand in U/V band for the uplink.}
Furthermore, we coordinate the lightweight core with a terrestrial 5G private network to test the integration of control \cite{wang2021}.
The architecture is shown in Figure 4.
In this figure, we show the log files of the lightweight 5G core network.
The lightweight 5G core is implemented in C and made up of three network functions: AMF, SMF, and UPF.
Based on that, the core network implements system procedures, including user registration and session establishment.
The onboard 5G core network interacts with the terrestrial 5G private network through control plane signaling.
The terrestrial 5G private network supports multiple access methods including fixed access and 5G new radio.
The private network is also able to offload traffic to the local edge network.

\begin{figure}
\centerline{\includegraphics[width=20pc]{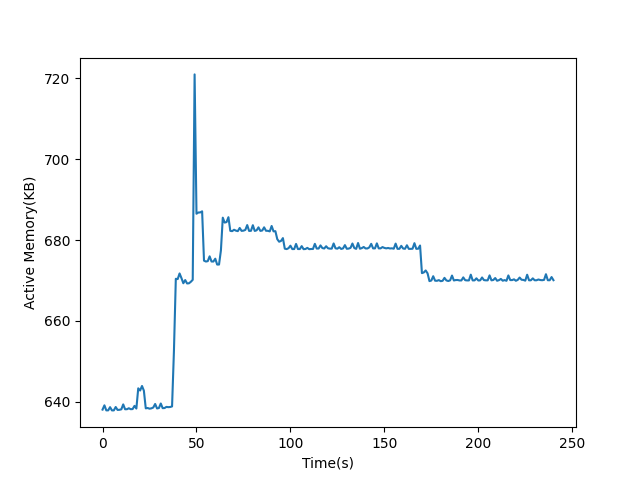}}
\caption{The active memory change over time.}
\end{figure} 

At first, we upload the lightweight 5G core network through the uplink operated at ultra high frequency. The satellite antenna receives the analog signal. The baseband decodes the signal and offloads the 5G core codebase to the onboard computer. The onboard computer then copy the code to a payload and executes it. Afterward, the telemetry shows that the 5G core network is deployed successfully on TY20. Figure 5 and Figure 6 show the CPU usage and active memory change.
The lightweight 5G core network starts at 38s and stops at 173s. The data is collected inside the onboard payload, Raspberry Pi 3 Model B+. We use the \textit{njmon} monitoring tool to observe the resource consumption. The system information is recorded per second. These results indicate that the 5G core network takes up 16\% of the CPU time and 20KB of memory. The log files shows that the three network functions work properly. The generated control plane signaling is also correct. Besides, the telemetry shows that the 5G core network can process signaling and data uploaded from the terrestrial 5G private network. The control plane signaling is downloaded to the terrestrial 5G private network. Finally, the 5G private network executes local traffic offloading and conducts business tests such as video calls.
In our test, we observe that the average latency of the signaling over satellite-terrestrial link is about 200 ms. Considering the wide-area network latency and processing latency, the result is acceptable for applications like video streaming or web browsing. We leave more comprehensive measurement study on the effectiveness of onboard 5G core networks as the future work.

\begin{table}[]
\centering
\caption{The Measurement on QUIC Procedures}
\label{tab:tb1}
\fontsize{6.5}{13}\selectfont
\begin{tabular}{@{}cccc@{}}
\toprule
Number & Pakcet Type & Elapsed Time(ms) & Length(byte)\\\midrule
1 & Initial & 0 & 1294\\
2 & Handshake & 2.95 & 1294\\
3 & Handshake & 4.93 & 1294\\
4 & Protected Payload & 5.83 & 1504\\\bottomrule
\end{tabular}
\end{table}

QUIC is a UDP-based multiplexed and secure transport protocol. It is considered the candidate for transport layer protocol of the next generation core network \cite{wang2022holistic}. To evaluate its feasibility, we experiment on an in-orbit satellite, TY22. We deploy a QUIC server on one of the payloads and a QUIC client on another. The two payloads link to the internal bus. Table 1 shows elapsed time and packet length during the QUIC communication process. The QUIC connection establishment is about 4.93 ms, which significantly outperforms the HTTP2 result (11.2ms on average) in our offline experiment.

\section{CONCLUSION}
{\setlength{\parindent}{0cm}
5G technology and satellite systems are complementary in function and demand.
The integration of the two will bring huge economic and social benefits.
As the control center of 5G systems, 5G core networks will play an important role in the satellite-terrestrial converged network.
This article looks forward to the cutting-edge development direction of the satellite-terrestrial integrated network.
It puts forward key points and a feasible architecture for the implementation of the 5G satellite core network.
Finally, we carried out the deployment and experimental verification of the lightweight 5G core network on an orbiting satellite.
According to the results, the satellite 5G core network can operate normally and perform the coordination between the satellite and the ground infrastructure.
Our experiment also serves as a reference to future research on B5G and 6G satellite-ground integrated communication networks.
}


\bibliographystyle{gbt7714-numerical}
\bibliography{myref}

\biographies
\begin{CCJNLbiography}{x.jpg}{Ruolin Xing}
received the bachelor degree in electronic information science and technology at Beijing University of Posts and Telecommunications of China in 2019. He is currently working toward the Ph.D. degree in computer science and engineering with the State Key Laboratory of Networking and Switching Technology, BUPT. His research interests include 5G/6G core network and satellite networks.
\end{CCJNLbiography}

\begin{CCJNLbiography}{mx.pdf}{Xiao Ma}
received her Ph.D. degree in Department of Computer Science and Technology from Tsinghua University, Beijing, China, in 2018. She is currently a lecturer at the State Key Laboratory of Networking and Switching Technology, BUPT. From October 2016 to April 2017, she visited the Department of Electrical and Computer Engineering, University of Waterloo, Canada. Her research interests include mobile cloud computing and mobile edge computing.
\end{CCJNLbiography}

\begin{CCJNLbiography}{za.pdf}{Ao Zhou}
received her Ph.D. degree in Beijing University of Posts and Telecommunications, Beijing, China, in 2015. She is currently an Associate Professor with State Key Laboratory of Networking and Switching Technology, Beijing University of Posts and Telecommunications. She has published 50+ research papers. She played a key role at many international conferences. Her research interests include cloud computing and mobile edge computing.
\end{CCJNLbiography}

\begin{CCJNLbiography}{Dustdar.png}{Schahram Dustdar}
is Full Professor of Computer Science with the Distributed Systems Group, TU Wien. He is an IEEE Fellow, an AAIA Fellow, and an elected member of the Academy of Europe, where he is the Chairman of the Informatics Section. He was a recipient of the ACM Distinguished Scientist Award, ACM Speaker award, IBM Faculty Award, IEEE TCSVC Outstanding Leadership Award for outstanding leadership in services computing. He is the Co-Editor-in-Chief of the ACM Transactions on Internet of Things and the Editor-in-Chief of Computing (Springer).
\end{CCJNLbiography}

\begin{CCJNLbiography}{wsg.pdf}{Shangguang Wang}
is a Professor at the School of Computer Science and Engineering, Beijing University of Posts and Telecommunications, China. He received his Ph.D. degree at Beijing University of Posts and Telecommunications in 2011. He has published more than 150 papers in journals such as IEEE JASC, TMC, TSC, and TCC, and conferences such as IJCAI, INFOCOM, AAAI and ICWS. His research interests include service computing, mobile edge computing, cloud computing, and Satellite Computing. He is currently serving as Chair of IEEE Technical Committee on Services Computing (2022-2013), and Vice-Chair of IEEE Technical Committee on Cloud Computing (2020-). He also served as General Chairs or Program Chairs of 10+ IEEE conferences, Advisor/Associate Editors of several journals such as Journal of Cloud Computing, Journal of Software: Practice and Experience, International Journal of Web and Grid Services, China Communications, and so on. He is a senior member of the IEEE.
\end{CCJNLbiography}

\clearpage

\end{document}